\documentclass{llncs}
\usepackage{makeidx}  
\usepackage{amsfonts}
\usepackage{amsmath}
\usepackage[dutch,USenglish]{babel}
\begin{document}

\title{A short note on the tractability of constructing phylogenetic networks 
from clusters}
\author{Leo van Iersel\inst{1}, Steven Kelk\inst{2}}
\institute{University of Canterbury, Department of Mathematics and Statistics,
Private Bag 4800, Christchurch, New Zealand, email: \texttt{l.j.j.v.iersel@gmail.com}. \and Centrum voor Wiskunde en Informatica 
(CWI), Life Sciences, P.O. Box 94079, 1090 GB Amsterdam, The Netherlands, email: \texttt{s.m.kelk@cwi.nl}.}

\maketitle              

\begin{abstract}
In \cite{cass} it was proven that the \textsc{Cass} algorithm is a 
polynomial-time algorithm for
constructing level-$\leq$2 networks from clusters\footnote{In this note we are referring exclusively to \emph{softwired} 
clusters, as opposed to hardwired clusters.}. Here we demonstrate, 
for each $k\geq 0$, a
polynomial-time algorithm for constructing level-$k$ phylogenetic networks 
from clusters. Unlike
\textsc{Cass} the algorithm scheme given here is only of 
theoretical interest. It does,
however, strengthen the hope that efficient polynomial-time algorithms 
(and perhaps fixed parameter tractable
algorithms) exist for this problem.
\end{abstract}
We refer the reader to \cite{cass} for definitions of the terminology used throughout this note.

\begin{theorem}
\label{thm:core}
Let $C$ be a set of clusters on a taxa set $X$ on $n$ leaves. Then, for every
fixed $k \geq 0$, it is possible to determine in polynomial time whether a level-$k$
network exists that represents $C$, and if so to construct such a network.
\end{theorem}
\begin{proof}
Suppose $C$ has the property that for every $X' \subset X$, $X'$ is
separated. We call such a cluster set $C$ \emph{fully separated}. 
It was shown in \cite{cass} that the existence
of a polynomial-time algorithm for constructing a level-$k$ network from a 
fully separated cluster set, is sufficient to give a polynomial-time algorithm for 
constructing level-$k$ networks from general cluster sets. (Specifically, the
fully-separated cluster sets are obtained by processing each non-trivial connected component of the 
\emph{incompatibility graph} of the original cluster set \cite{cass}).
Hence we can assume
without loss of generality that $C$ is fully separated. Furthermore, it was
also shown in \cite{cass} that (i) any network that represents $C$ is simple, and (ii)
if there exists a level-$k$ network that represents $C$, then there exists 
a binary simple level-$k$ network $N$ that represents $C$ \cite{cass}. Lemma
\ref{lem:core} is thus sufficient, and we are done. \qed
\end{proof}

\begin{lemma}
\label{lem:core}
Let $C$ be a fully separated cluster set on a taxa set $X$, where $|X|=n$. Then, for every fixed
$k \geq 0$, it is possible to determine in polynomial time whether a binary simple level-$k$ network
exists that represents $C$, and if so to construct such a network.
\end{lemma}
\begin{proof}
We assume that $k$ is fixed. Assume then that there does exist a binary simple level-$k$ network
$N$ that represents $C$. Then $C$ will contain at most $2^{k+1}(n-1)$ clusters, 
because there are at most $2^{k}$ trees displayed by a binary simple 
level-$k$ network, and each tree represents at
most $2(n-1)$ clusters. Thus, for fixed $k$, the size of the input is 
polynomial in $n$. Note also that because of these facts it is easy to 
check in polynomial time whether a set of clusters is indeed represented by a 
given simple level-$k$ network. In the remainder of this proof we will
use this fact implicitly to distinguish correct from incorrect guesses.

It is known that, if the leaves of $N$ are removed and all vertices with both 
indegree and outdegree
equal to 1 are suppressed, the resulting structure will
be a level-$k$ generator, defined in \cite{lev2}. For fixed $k$, there are only 
a constant
number of level-$k$ generators. Recall that the \emph{sides} of a 
level-$k$ generator are
defined as the union of its edges and its vertices of indegree-2 and 
outdegree-0. For fixed $k$ the maximum number of sides ranging
over all level-$k$ generators, is a constant.\\
\\
For a cluster set $C$ on $X$, we write $x \rightarrow y$ iff every 
non-singleton cluster in $C$ that contains $x$, also contains $y$.
Let $G(C) = (X,E)$ be a directed graph on $X$ with edge set as follows:
$(x,y) \in E$ iff
$x \rightarrow y$ and there is no $z \not \in \{x,y\}$ such that $x 
\rightarrow z$ and $z \rightarrow y$. It is easy to see
that $G(C)$, which we call the \emph{containment graph}, is acylic: the 
presence of cycle in $G(C)$ would
mean that all subsets of the taxa in the cycle are unseparated, contradicting the 
assumption of the lemma.\\
\\
We propose the following simple algorithm for determining whether $C$ is 
represented by a binary simple level-$k$ network. In particular, we
will attempt to reconstruct $N$. Let $g$ be the generator underlying $N$. 
We only require polynomially many ``guesses'' to
compute $g$, because there are only a constant number of generators. So 
assume we know $g$. For each side of $g$, we guess whether there
are 0, 1, 2 or more than 2 leaves on that side. For each side containing 
exactly one leaf, we guess what that is. For each side $s$
of $g$ containing 2 or more leaves, we guess the leaf $s^{+}$ that
is nearest to the root on that side, and the leaf $s^{-}$ that is
furthest from the root on that side.

We will now show how to add the remaining leaves. The critical point to 
note is that every remaining leaf will be added between the $+$ and the 
$-$ leaf on some side.  We add the remaining leaves in a specific order. 
In particular, we say that a side $s$ is \emph{lowest} if it does not yet 
have all its
leaves, and there is no other such side $s'$ reachable from $s$. By 
reachable we mean: in the underlying generator $g$, there is a directed 
path from the head of side $s$ to the tail of side $s'$. 
(The sides for which we guessed that they have 0, 1 or exactly 2
leaves, can never be lowest). $N$ is a directed acyclic graph, so
until all remaining leaves have been added, there will always be
a lowest side.

The idea is to add leaves to the lowest side $s$, until all its leaves 
have been added. We then continue with remaining lowest
sides until we have reconstructed $N$.

It is possible to tell in polynomial time what the correct leaves are for 
that side, as follows. Observe that a leaf $x$ that is on side
$s$ in $N$ has the property $s^{+} \rightarrow x \rightarrow s^{-}$.
(Clearly $x \not \rightarrow s^{+}$ and $s^{-} \not \rightarrow x$).
Furthermore, there is at least one cluster $c \in C$ such that
$\{ x, s^{+}, s^{-} \} \cap c = \{ x, s^{-} \}$. We call such a
cluster a \emph{split cluster for side $s$}. There exists at least
one such cluster because otherwise $\{ x, s^{+} \}$ would be unseparated 
in $C$. Now, observe that for every split cluster $c$
for side $s$, and for every side $t \neq s$ that contains 2 or more leaves 
in $N$, either $\{t^{+}, t^{-}\} \cap c = \{t^{+}, t^{-}\}$ or
$\{t^{+}, t^{-}\} \cap c = \emptyset$. This follows because the
only edges in $N$ that represent $c$ lie on side $s$. Now, consider
any leaf $y$ that has not yet been added to the network and is not
on side $s$ in $N$, but side $t$ (for some $t$). Side $t$ will
contain three or more leaves in $N$, so we can assume that $t^{+}$
and $t^{-}$ exist. If it is not the case that
$s^{+} \rightarrow y \rightarrow s^{-}$ then it is immediately
clear that $y$ cannot be put on side $s$. So assume (conversely)
this condition does hold, and for the same reason assume there is a split 
cluster $c$ for side $s$ that contains $y$. In other words,
there is a cluster $c$ such that $\{ y, s^{+}, s^{-} \} \cap c = \{ y, 
s^{-} \}$. It follows that $c$ also contains $t^{+}$ and $t^{-}$,
because (by inspection on $N$) any cluster that contains $y$ also
contains $t^{-}$, and we know that $c$ contains either both of
$t^{+}$ and $t^{-}$, or neither of them. However, there is no edge
in $N$ that can represent $c$: the only edges that represent $c$
lie on side $s$, but the fact that $s$ is the lowest side means that
no cluster beginning on side $s$ can contain any leaves on side $t$.
To summarise, then, we have a simple test for determining whether
a leaf should be placed on side $s$. Once we have determined the set of 
leaves that should be placed on side $s$, it is easy to determine
the correct order of those leaves by inspecting the containment
graph. This concludes the proof. \qed
\end{proof}
It is interesting to note that the above proof technique leads to a simplified
proof, presented in the following Corollary, of a result that was first
proven in \cite{simplicity}. (The algorithm presented in \cite{simplicity} was, however,
far more efficient). We refer the reader to \cite{lev2}\cite{simplicity} for definitions related to triplets.

\begin{corollary}
Let $T$ be a dense set of triplets on leaf set $L$ on $n$ leaves. Then, for every fixed
$k \geq 0$, it is possible to determine in polynomial time whether a binary simple level-$k$ network
exists that is consistent with $T$, and if so to construct such a network.
\end{corollary}
\begin{proof}
The proof of Lemma \ref{lem:core} holds here almost entirely. (As noted in \cite{simplicity} it is possible to
determine in polynomial time whether a given network is indeed consistent with a set of input triplets). The only
significant difference concerns the adding of leaves to the lowest side. The crucial fact here is that a not yet allocated leaf $x$ belongs on 
lowest side $s$ if and only if the triplet $s^{-}x|s^{+}$ is in the input. \qed
\end{proof}



\begin{thebibliography}{77}	

\bibitem{lev2} L. J. J. van Iersel, J. C. M. Keijsper, S. M. Kelk, L. Stougie, F. Hagen, and T. Boekhout, Constructing level-2
phylogenetic networks from triplets, \emph{IEEE/ACM Transactions on Computational Biology and Bioinformatics} 6(4),
pp. 667--681, October 2009.

\bibitem{cass} L. J. J. van Iersel, S. M. Kelk, R. Rupp and D. Huson,
Phylogenetic Networks Do not Need to Be Complex: Using Fewer Reticulations to Represent Conflicting Clusters,
arXiv:0910.3082v1 [q-bio.PE], October 2009.

\bibitem{simplicity}  L.~J.~J. van Iersel and S.~M. Kelk. 
\newblock Constructing the simplest possible phylogenetic network from
  triplets,
\newblock {\em Algorithmica}, 2009.
\newblock To appear.


\end{thebibliography}
\end{document}